\documentclass[fleqn,usenatbib]{mnras}

\usepackage{newtxtext,newtxmath}

\usepackage[T1]{fontenc}
\usepackage{ae,aecompl}

\usepackage{graphicx}	
\usepackage{amsmath}	
\usepackage{amssymb}	

\title[Star formation in high z galaxies]{UV regulated star formation in high-redshift galaxies}

\author[Latif \& Khochfar]{
Muhammad A. Latif,$^{1}$\thanks{E-mail: latifne@gemail.com}
Sadegh Khochfar$^{2}$ \\
$^{1}$Physics Department, College of Science, United Arab Emirates University, PO Box 15551, Al-Ain, UAE \\
$^{2}$Institute for Astronomy, University of Edinburgh, Royal Observatory, Blackford Hill, Edinburgh EH9 3HJ, UK}

\date{Accepted XXX. Received YYY; in original form ZZZ}

\pubyear{2019}

\begin{document}
\bibliographystyle{mnras}

\label{firstpage}
\pagerange{\pageref{firstpage}--\pageref{lastpage}}
\maketitle

\begin{abstract}
The first galaxies forming a few hundred million years after the Big Bang are the key drivers of cosmic evolution and ideal laboratories to study theories of galaxy formation. We here study the role of UV radiation in suppressing star formation in primordial galaxies by destroying molecular hydrogen, the main coolant in primordial gas and provide estimates of cold dense gas at the onset of star formation. To accomplish this goal, we perform three dimensional cosmological simulations of minihalos in different environments  forming at $ z\sim 25$ by varying strength  of background UV flux below the Lyman limit  between 0.01-1000 in units of $\rm J_{21}=10^{-21}~erg/cm^2/s/Hz/sr$. Particularly, we include photo-detachment of $\rm H^-$, the self-shielding of $\rm H_2$ which both were  neglected  in previous studies and use updated reaction rates.  
Our results  show that depending on the background level $\rm H_2$ formation is suppressed, delaying gravitational collapse until halos reach the atomic cooling limit. We find that the formation of cold dense molecular gas and subsequently star formation gets delayed by 100 to 230 Myr depending on  the level of the background radiation and the growth history of the dark matter halos. 
The fraction of dense self-shielded gas  is a strong function of the background flux and exponentially declines with the strength of incident UV flux above $\rm J_{21} \geq 1$. We find that taking into account  $\rm H_2$ self-shielding  is crucial for accurately estimating the amount of cold dense gas available for star formation. 


\end{abstract}

\begin{keywords}
methods: numerical -- cosmology: theory -- early Universe -- high redshift-- galaxies formation-- radiation
\end{keywords}



\section{Introduction}
Remarkable progress in high-redshift surveys have pushed the observational frontier up to the cosmic dawn. More than 800 galaxies have been detected  within the first billion years after the Big Bang  with candidate galaxies up to $z \sim 11$\citep{Bouwens16,Oesch16,Mcleod2016,Lam19}. These galaxies are the potential hosts of the first and second generation of stars and key laboratories to test theories of  galaxy formation. When and how these galaxies formed stars are key questions in the study of high-redshift galaxy formation.

Our understanding of structure formation is based on the $\Lambda \rm{CDM}$ paradigm. According to the hierarchical scenario of structure formation, the first generation of stars so-called Pop III stars are formed in minihalos  at $z \sim 30$. First studies estimated that Pop III stars are more massive than contemporary stars in the local Universe with masses about a few hundred solar masses \citep{Abel2000, Bromm2002,Yoshida2008}. However, recent high resolution numerical simulations suggest that proto-stellar discs form during the collapse of gas with non-negligible angular momentum, which can become unstable due to gravitational torques and fragment to form multiple stars per minihalo \citep{Turk09, Clark11,Greif12,Latif2013ApJ,Hirano2014, Stacy2016}. Consequently, the typical masses of Pop III stars are expected to be about a few tens of solar masses and lower than in previous studies. Nevertheless, some of the clumps formed due to disc fragmentation may migrate inwards and merge with the central protostar resulting in an intermittent UV feedback from the star \citep{Latif2015Disk, Hosokawa16}. Pop III stars, depending on their mass, are expected to influence the subsequent star formation via their chemical  \citep{Bromm01,Bromm03,Schneider2003,Omukai2005,Smith07}, mechanical \citep{MacLow99,Nishi99,Bromm03,Springel03,Mackey03} and radiative feedback \citep{Haiman97,Haiman00,Yoshida2003,Ciardi2005,Wise2008A,Johnson08,Johnson13}.

Pop III stars are expected to form out of a primordial gas in minihalos of $\rm 10^5-10^6~M_{\odot}$ with virial temperatures of about a few thousand K. In primordial halos, a trace amount of molecular hydrogen can be formed via gas-phase reactions which can cool the gas down to about 200 K. In the absence of dust and metals, the gravitational collapse in these minihalos is induced by molecular hydrogen. The first generation of stars are expected to have short lifetimes depending upon their mass scale \citep{Heger2002,Heger2003}, they may go off as supernovae and enrich the intergalactic medium with metals \citep{Maio2011,Smith2015}. The second generation of stars so-called Pop II stars are formed from  metal enriched gas with metallicity as low as $\rm Z/Z_{\odot} \sim 3 \times 10^{-4}$ \citep{Schneider2003,Omukai2005,GloverJappsen2007,Wise2012a,Bovino14,Latif2016dust}. The value of the critical metallicity may further get reduced by an order  of magnitude in the presence of dust cooling.  Pop II stars are expected to be more abundant and also have longer lifetimes compared to the Pop III stars due to their lower masses on average.

The first galaxies hosting Pop II and Pop III stars are presumed to emit copious amount of UV radiation and may influence the star formation in nearby or even in distant halos \citep{Johnson13}. While UV radiation above 13.6 eV gets absorbed in the surrounding gas, UV photons below the Lyman limit (13.6 eV) have long mean free paths and may regulate the star formation by suppressing the amount of cold dense  gas in primordial halos. The UV radiation below the Lyman limit can photo-detach $\rm H^-$ which is the main pathway for $\rm H_2$ formation as well as can directly dissociate $\rm H_2$ molecules via the Solomon process. \cite{Machacek01} performed numerical simulations to study the impact of Lyman Werner (11.2-13.6 eV) radiation on primordial gas clouds. They found that such soft-UV radiation can delay star formation and that the amount of cold gas depends on the strength of LW flux as well as cloud mass. However, they ignored the effects of $\rm H_2$ self-shielding and $\rm H^-$ photo-detachment in their study. Both of these factors are important in estimating the cold dense gas  available for star formation in primordial halos. In addition to that a number of updates to reaction rates for $\rm H^-$ detachment, $\rm H_2$ photo-dissociation and three-body reactions have been introduced, which all affect the amount of cold dense star forming gas in halos.  Therefore, it is worthwhile to revisit this problem with state of the art chemical models and higher resolution simulations.

To assess the role of UV radiation in regulating the star formation in high-redshift galaxies, we perform a suite of 3D cosmological simulations for 6 distinct halos with different merger histories by employing a UV flux with $\rm T_{rad}= 2 \times 10^4 ~K$ which can mimic realistic spectra of the first galaxies \citep{Agarwal2015,Sugimura14, Latif2015a}. We employ  state of the art chemical models with the latest reaction rates \citep{Latif2015a, Glover2015a}. We vary the strength of UV field from 0.01 -1000 in units of 
$ \rm J_{21} = 10^{-21} erg ~cm^{-2} ~s^{-1} ~Hz^{-1} ~sr^{-1}$. Our findings suggest that the amount of star forming cold gas decreases with UV flux and collapse gets significantly delayed. Our work provides new estimates of cold dense star forming gas for a given strength of UV background radiation below the Lyman limit.

This article is organized in the following way. In section 2, we provide details of numerical methods and chemical model employed in this work. In section 3, we present our main findings and confer our conclusions in section 4.

\section{Numerical methods}
We use the open source code Enzo to carry out cosmological hydro-dynamical simulations \citep{Enzo2014}. Enzo is an adaptive mesh refinement (AMR), parallel, grid based, Eulerian code which can run on different platforms and is suitable for cosmological simulations. We employ the piece-wise parabolic method (PPM) to  solve hydrodynamical equations and make use of particle-mesh based N-body solver to compute dark matter (DM) dynamics. We use multigrid Poisson solver for self-gravity calculations.

Our simulations are commenced with cosmological initial conditions generated from the MUSIC package \citep{Hahn2011} at $z=150$ using the Planck 2016 data with $\Omega_{\rm M}=0.3089$, $\Omega_{\Lambda}=0.6911$, $\rm H_{0}=0.6774$ \citep{Planck2016}. They have top grid resolution of $256^3$ and the same number of DM particles in a cosmological volume of $\rm 1 Mpc/h$. We further employ two additional nested grids each with the same resolution ($256^3$ grid cells) as of the top grid, yielding an effective resolution of $1024^3$. Consequently, the maximum DM resolution in our simulations is $\rm 67~ M_{\odot}/h$. We subsequently employ further 20 levels of dynamical refinement during the course of simulations by exploiting the AMR capability which yields spatial resolution of about 50 AU in proper units. Our refinement criteria is based on  baryonic overdensity, Jeans refinement and the DM mass. A cell is flagged for refinement once it exceeds four times the cosmic mean density or DM particle density of 0.0625 times $\rho_{\rm{DM}}r^{\ell \alpha}$ where $\rho_{\rm{DM}}$ is the dark matter density, $r = 2$ is the refinement factor, $\ell$ is the refinement level, and $\alpha = -0.3$ makes the refinement super-Lagrangian. The Jeans length in our simulations is resolved  at-least by 32 cells throughout their evolution. We selected 6 distinct halos with different merger histories, various environments and are placed at the center of the box.

We employ the publicly available astrochemistry package KROME \citep{Grassi2014} to solve the  chemical and thermal evolution of  primordial gas chemistry along with cosmological simulations. In our model we solve the rate equations of the following nine primordial species $ \rm H, ~H^+,~ H^-, ~He,~ He^+, ~He^{++},~ H_2, ~H_2^+, ~e^-$ and  is based on \cite{Latif2015a}, all reaction rates and cooling/heating functions are listed in their table 1 in appendix A.  Photons with energy range from 11.2-13.6 eV so-called Lyman-Werner photons can directly dissociate molecular hydrogen while photons with energy  greater than 0.76 eV can photo-detach $\rm H^-$ which is main pathway for $\rm H_2 $ formation. To cover both  ranges our model mimics  the full relevant spectral energy distribution  below the Lyman limit. The choice of our spectra are consistent with those of the first galaxies as shown by e.g. \cite{Sugimura14,Agarwal2015,Latif2015a}. The intensity of  simulated  UV flux ranges  from 0.01-1000 in units of $\rm J_{21}=10^{-21}~erg/cm^2/s/Hz/Sr$ and is assumed to be generated by both the global background as well as local flux emitted from nearby halos. We assume that UV radiation below the Lyman limit are  emitted by the first galaxies with black body radiation temperature of $\rm 2 \times 10^4$ K\footnote{see \cite{Agarwal19} for deviations from this assumption and the impact on reaction rates}.  Apart from photo-detachment of $\rm H^-$ and photo-dissociation  of $\rm H_2$, we also include collisional-dissociation of $\rm H_2$ and $\rm H_2^+$.  Our model includes cooling and heating from all the relevant processes such as cooling from collisional excitation, collisional ionization, radiative recombination, collision induced emission, bremsstrahlung radiation as well as $\rm H_2$ cooling, chemical heating/cooling from three-body  reactions. For $\rm H_2$ self-shielding we use the fitting formula by \cite{WolocottGreen2011}. Further details of our chemical model are discussed in \cite{Latif2015a} and \cite{Latif2016}. We have ignored the Deuterium related processes as HD gets dissociated even in the ubiquity of a very weak UV flux. Therefore, we expect that it will have no influence on our findings.

\begin{table*}
\begin{center}
\caption{Virial masses and the collapse redshifts of simulated halos are listed here against the the strength of UV flux. We mention the average cold dense gas fraction in the last column.}
\begin{tabular}{| c | c | c | c |c | c| |c| c|c|}
\hline
\hline
 
UV flux   & Halo 1 & Halo 2 & Halo 3 & Halo 4 & Halo 5 & Halo 6 & $\rm f_{cd}$\\

$\rm J_{21}$ & $z$, $\rm M_{\odot}$  &$z$, $\rm M_{\odot}$ &$z$, $\rm M_{\odot}$ &$z$, $\rm M_{\odot}$ &$z$, $\rm M_{\odot}$ &$z$, $\rm M_{\odot}$ & mean value  \\
\hline                                                          \\
0.1 & 24.5, $\rm 8.8 \times 10^{5}$   & 24.7, $\rm 8.0 \times 10^{5}$  & 23.6, $\rm 1.1 \times 10^{6}$  & 25.6, $\rm 4.2 \times 10^{5}$  & 24.3, $\rm 6.3 \times 10^{5}$ & 25.4, $\rm 7.6 \times 10^{5}$ & 0.0017 \\
1   & 23.9, $\rm 8.0  \times 10^{5}$  & 22.5,$\rm 1.4  \times 10^{6}$  & 22.3, $\rm 2.0  \times 10^{6}$  &21.9,$\rm 1.0  \times 10^{6}$ & 20.1, $\rm 2.6  \times 10^{6}$ & 24.5,$\rm 1.2  \times 10^{6}$ & 0.0012\\
5   &15.4, $\rm 7.7 \times 10^{6}$   & 21.2,$\rm 2.5 \times 10^{6}$  & 20, $\rm 9.4 \times 10^{6}$   & 14.2,$\rm 7.8 \times 10^{6}$  &16.4, $\rm 1.0 \times 10^{7}$ & 13.3,$\rm 1.8 \times 10^{7}$ &0.00014 \\
10  &15, $\rm 9.3 \times 10^{6}$  & 13.5,$\rm 2.4 \times 10^{7}$  &19,$\rm 1.2 \times 10^{7}$  &14,$\rm 9.4 \times 10^{6}$  &14.5,$\rm 2.5 \times 10^{7}$  &11.5,$\rm 5.7 \times 10^{7}$  &0.00015 \\
50  &12.6,$\rm 1.5 \times 10^{7}$  & 13, $\rm 2.6 \times 10^{7}$ &18,$\rm 1.5 \times 10^{7}$  &12,$\rm 9.4 \times 10^{7}$ & 14.3,$\rm 2.8 \times 10^{7}$ & 11.6,$\rm 5.0 \times 10^{7}$ & $\rm 9.8 \times 10^{-5}$ \\
100 & 11.7,$\rm 1.8 \times 10^{7}$  &13,$\rm 2.6 \times 10^{7}$ & 18,$\rm 1.5 \times 10^{7}$ &11.8,$\rm 9.4 \times 10^{7}$ & 13.5,$\rm 4.0 \times 10^{7}$ & 11,$\rm 6.3 \times 10^{7}$ &$\rm 8.0 \times 10^{-5}$ \\
1000 &11.4, $\rm 1.9 \times 10^{7}$  & 12.3,$\rm 5.3 \times 10^{7}$ & 17,$\rm 1.8 \times 10^{7}$   & 11,$\rm 1.96 \times 10^{8}$ &13.2,$\rm 4.2 \times 10^{7}$ &10.6,$\rm 6.7 \times 10^{7}$ &$\rm 3.0 \times 10^{-5}$ \\
\hline
\end{tabular}
\label{table1}
\end{center}
\end{table*}

\section{Results}
\subsection{Overview}
We present our main findings in this section. In total, we have performed about 50 cosmological simulations for six different halos turning on the UV background at $\rm z = 30$ and varying the strength of the impinging UV radiation field from 0.01-1000 in units of $\rm J_{21}$. The simulated halos have different merger histories and are  selected from environments with different dark matter over-densities ranging from 0.1 to 100 times the cosmic mean in a volume of 1(Mpc/h)$^3$ centred around the halo of interest.  The mass range of selected halos varies from a few times $\rm 10^5~M_{\odot}$ to $\rm 10^6~M_{\odot}$ and they typically form at $z=23-26$. We stop our simulations soon after they reach 20 levels of refinement which gives us a spatial resolution of about $\rm 10^{-3}$ pc in physical units and gas densities of about $\rm 10^{-15}~g/cm^3$ ($\rm 10^9 ~cm^{-3}$). In the next subsections, we discuss physical properties of gas in halos and compute the mass of $\rm H_2$ cooled dense gas, followed by estimates of cold gas fractions available for star formation given the strength of the UV background. 

\subsection{Halo gas properties}

\begin{figure*} 
\begin{center}
\includegraphics[scale=0.8]{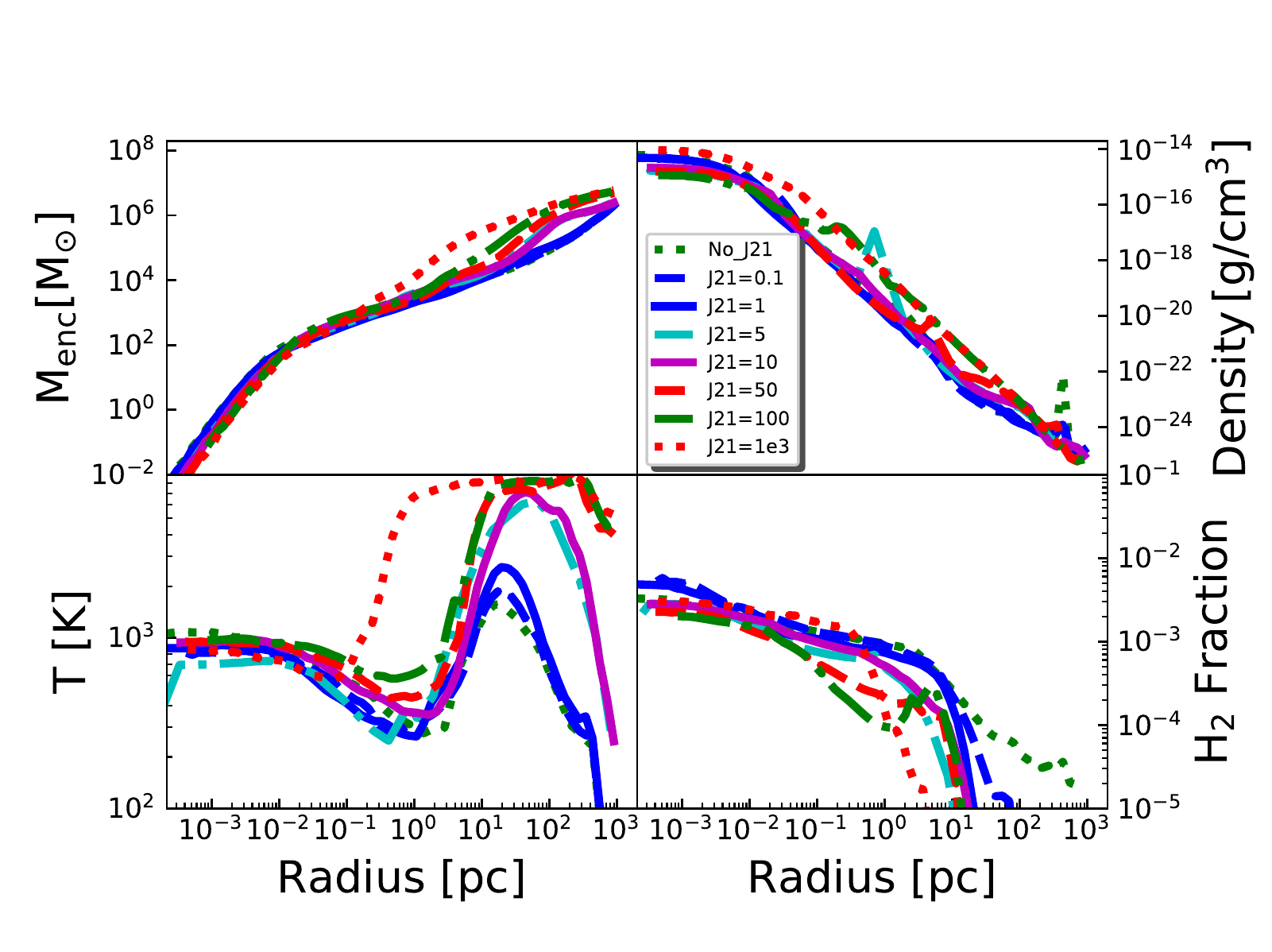}
\end{center}
\caption{Spherically averages and radially binned profiles of gas density, enclosed mass, temperature and $\rm H_2$ fraction. Different line styles and colors represent the strength of the UV field as mentioned in the legend. This plot is made for  halo 1 and the collapse redshifts  for corresponding $\rm J_{21}$ are listed in table 1.}
\label{fig}
\end{figure*}

\begin{figure*} 
\begin{center}
\includegraphics[scale=0.8]{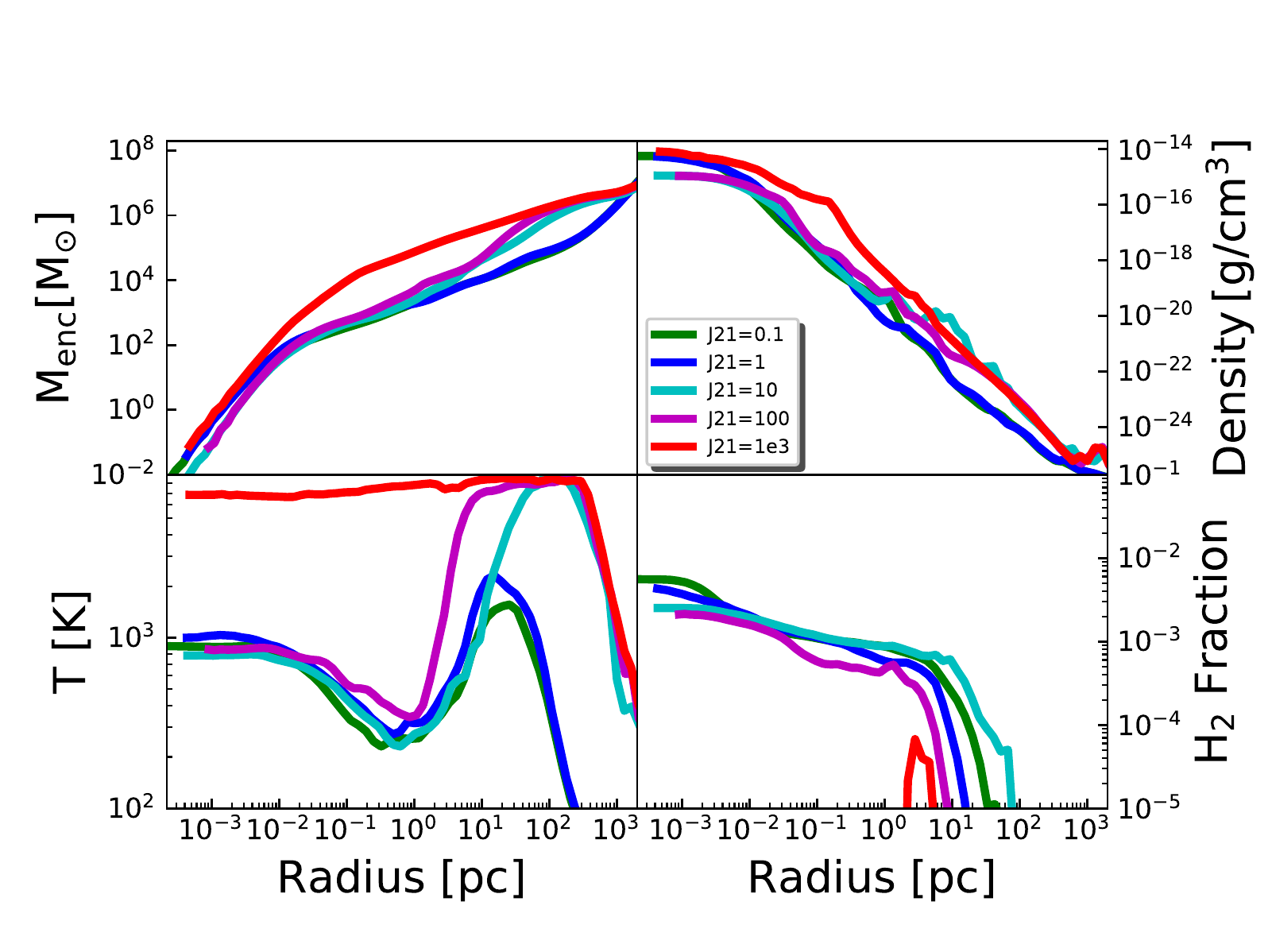}
\end{center}
\caption{Spherically averages and radially binned profiles of gas density, enclosed mass, temperature and $\rm H_2$ fraction. Different line styles and colors represent the strength of UV field as mentioned in the legend. This plot is made for  halo 6 and the collapse redshifts  for corresponding $\rm J_{21}$ are listed in table 1.}
\label{fig1}
\end{figure*}

Gas initially falls into the DM potentials and gets heated via shocks and mergers until it reaches virial temperatures  of about a few thousand Kelvin. Virialization shocks further catalyze $\rm H_2$ formation and consequently the $\rm H_2$ fraction reaches  above $\rm 10^{-3}$. In the absence of a UV radiation flux, collapse gets triggered by molecular hydrogen in a halo of a few times $\rm 10^5~M_{\odot}$ and the gas temperature decreases down to a few hundred K. The $\rm H_2$ fraction further gets boosted in the center of the halo via three-body reactions at densities above $n \geq 10^6 ~{\rm cm}^{-3}$. For a weaker radiation field of strength $\rm  J_{21} \leq 0.1$, the impact of UV radiation is almost negligible as the collapse is slightly delayed but the fraction of molecular hydrogen in the halo is comparable to the no UV case. Consequently, the thermal evolution is similar to the no UV case mainly due to the effect of $\rm H_2$ self-shielding, see Fig. \ref{fig} and Fig. \ref{fig1} for two representative halos.  This is true for all simulated halos.  

For $\rm J_{21} > 1$, UV radiation suppress $\rm H_2$ formation  via $\rm H^-$ photo-detachment as well as directly dissociate $\rm H_2$ molecules via the Solomon process. Due to the lack of sufficient $\rm H_2$, the collapse gets delayed until the halo mass reaches the  atomic cooling limit at which point atomic line cooling kicks in and $\rm H_2$ self-shielding from UV radiation becomes effective in the center of the halos as $\rm H_2$ column density exceeds  $\rm  10^{14}~cm^{-2}$. Consequently, the $\rm H_2$ fraction gets boosted and cools the gas in the core of the halos down to a few hundred K similar to the weaker $\rm J_{21}$ cases. At higher densities above $\rm 10^{-18}~ g/cm^3$, the gas temperature starts to increase  due to the compressional heating and also the gas becomes optically thick to $\rm H_2$ line cooling. The same trend is observed for all simulated halos. The average density profile follows a $\rm r^{-2.1}$ behaviour as expected from primordial gas collapse \citep{Greif12, Latif2013ApJ} and profiles become flat within the centre of halos which corresponds to the central Jeans length. The bumps on the density profile indicate the presence of small substructure in the halo. The mass profile sharply increases in the central core corresponding to the central Jeans length and then almost linearly increases with radius following the density profile. The enclosed mass is enhanced for stronger $\rm J_{21}$ cases due to the larger halo mass. Similar behaviour is found for all halos.

\begin{figure*} 
\begin{center}
\includegraphics[scale=0.8]{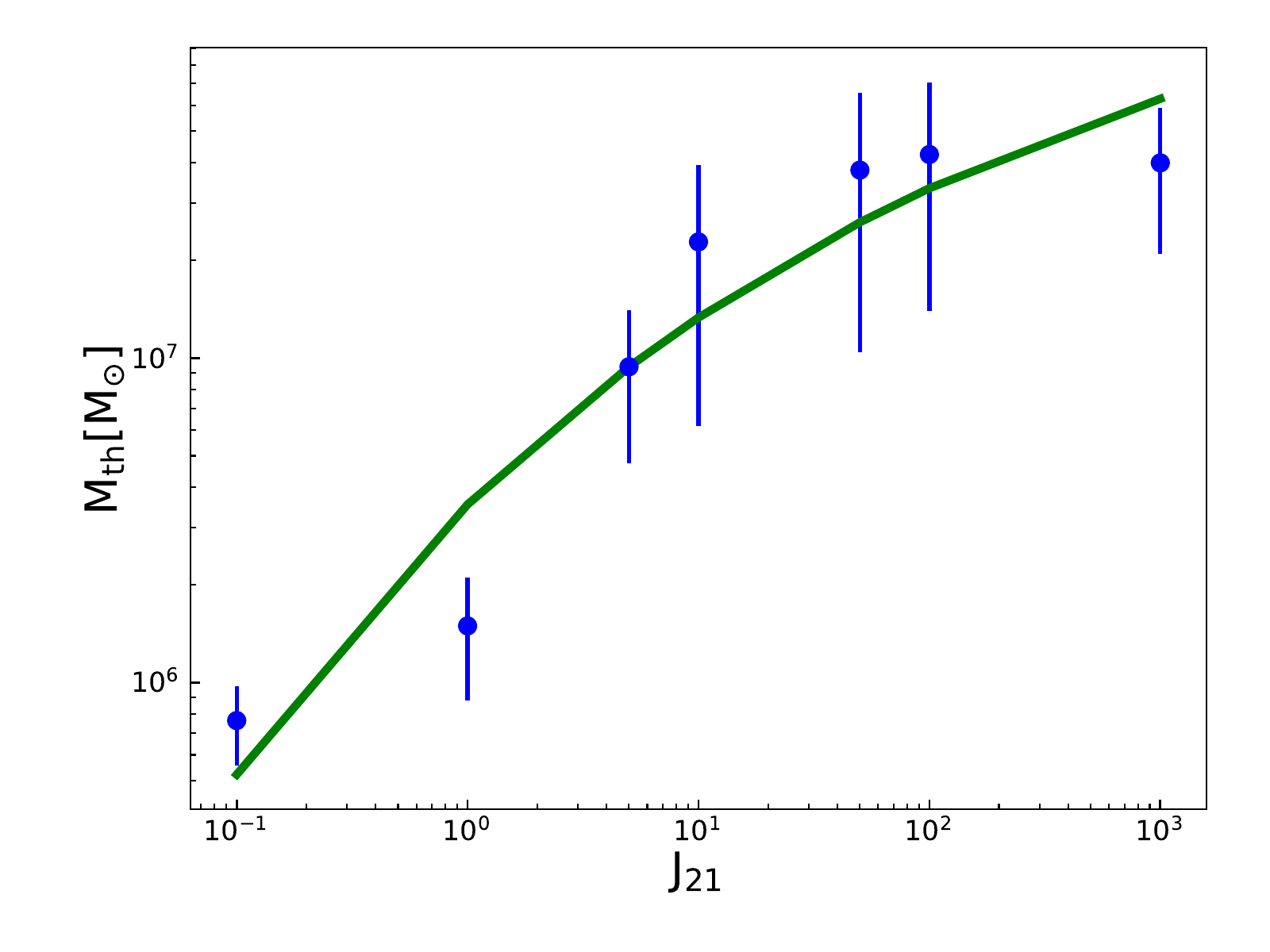}
\end{center}
\caption{The critical halo mass is plotted against the strength of UV flux in units of $\rm J_{21}$. The error-bars indicate the 1-sigma dispersion in the  mass for simulated halos. The green solid line is fit to the data given in Eq. 1. }
\label{fig2}
\end{figure*}

\begin{figure*} 
\begin{center}
\includegraphics[scale=0.8]{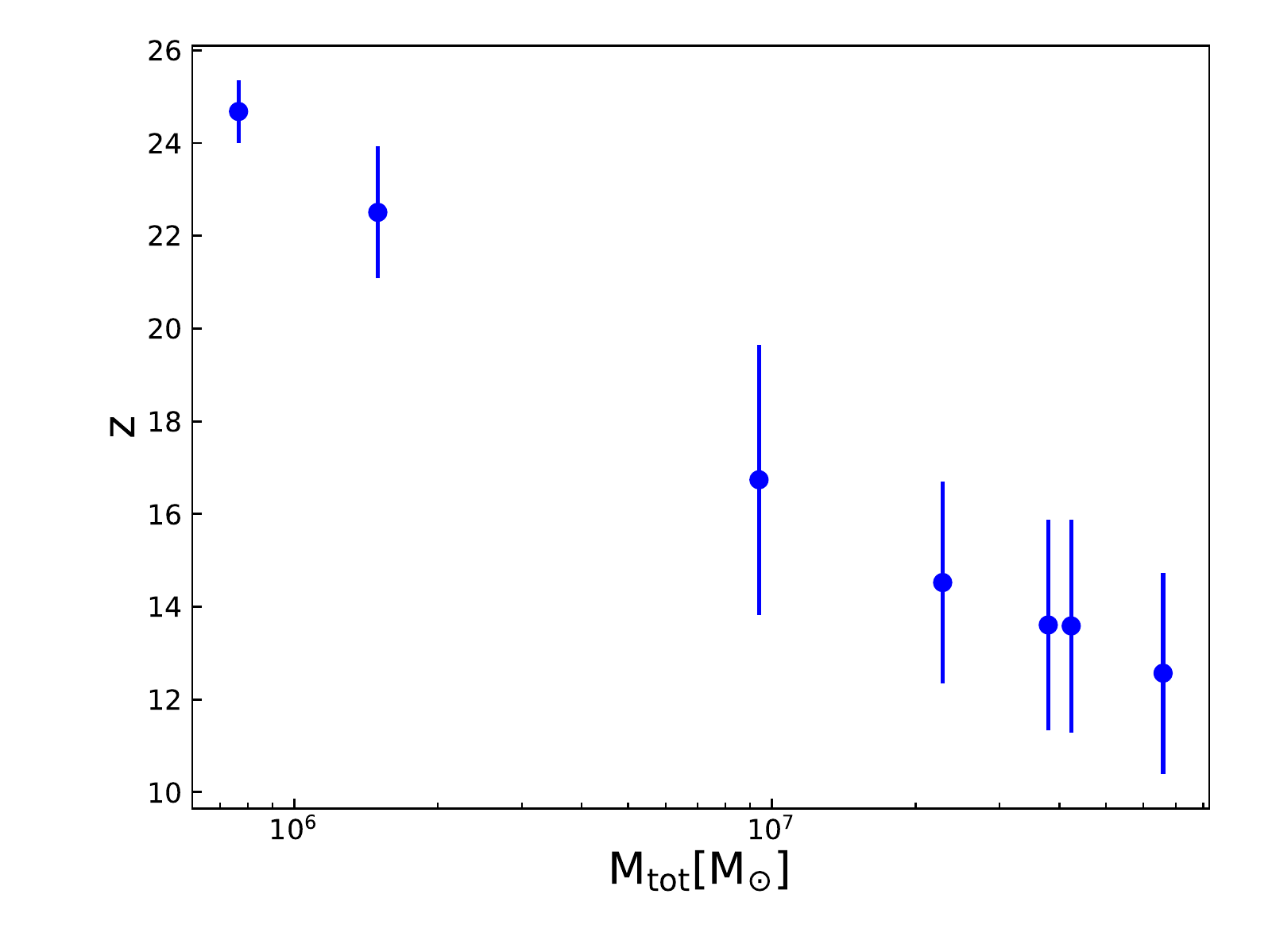}
\end{center}
\caption{ This plot shows the collapse redshift against the halo mass for simulated strengths of UV flux ranging from 0.1-1000 in units of $\rm J_{21}$ and error-bars indicate 1-sigma dispersion in the collapse redshifts of simulated halos.}
\label{fig3}
\end{figure*}

Interestingly, in one of the simulated halos (halo 6), the collapse proceeds isothermally for $\rm J_{21}=1000$  and the $\rm H_2$ fraction  remains as low as $\rm 10^{-7}$. This comes from the fact that halo 6 has a quiescent growth history and has not gone through any major merger. This results in a different density structure in halo 6 reducing the ability of gas to effectively self-shield $\rm H_2$ compared to halo 1. Previous studies exploring the critical value of UV flux required for an isothermal collapse have shown that it varies from halo to halo and depends on the local density distribution which is affected e.g. by the merger history and collapse redshift (see \cite{Latif2014UV} and \cite{Latif2015a} for a detailed discussion). Under such conditions, we expect that halo 6  will directly collapse into a massive black hole instead of forming ordinary stars. The minimum  halo threshold mass for gas collapse and the onset of star formation increases with the strength of UV flux,  it reaches  $\rm 10^7~M_{\odot}$ for $\rm J_{21} = 10$ and continues to increase for stronger radiation fields until $\rm 5 \times 10^7~M_{\odot}$, the atomic cooling limit, see Fig. \ref{fig2}. Overall the halo threshold mass becomes almost constant for $\rm J_{21} \geq 50$ with some scatter due to the variations in the halo growth history. Similarly, the collapse of gas in halos  gets delayed in the ubiquity of a UV flux.  This comes from the fact that $\rm H_2$ suppression becomes more effective with enhancing the intensity of UV radiation, consequently gas cannot cool until the halo virial temperature reaches $\rm \geq 10^4~K$ and atomic line cooling kicks in. This results in longer collapse time scales,  delayed cold dense molecular gas formation and subsequently late star formation depending upon the UV field strength. For stronger radiation fluxes ($\rm J_{21}=10$), the gas collapse in halos gets delayed up to $\rm \Delta z \sim 14$ (about 230 Myrs) and the critical halo mass  increases by almost two orders of magnitude as shown in Fig. \ref{fig3}. On average for $\rm J_{21}> 5$ the collapse is delayed by about $\rm 100~Myrs$. The error-bars in halo threshold mass  and collapse redshifts correspond to one sigma value and the dispersion is due to different halo merger and growth histories. 

\subsection{Cold gas fraction}
\begin{figure*} 
\begin{center}
\includegraphics[scale=0.8]{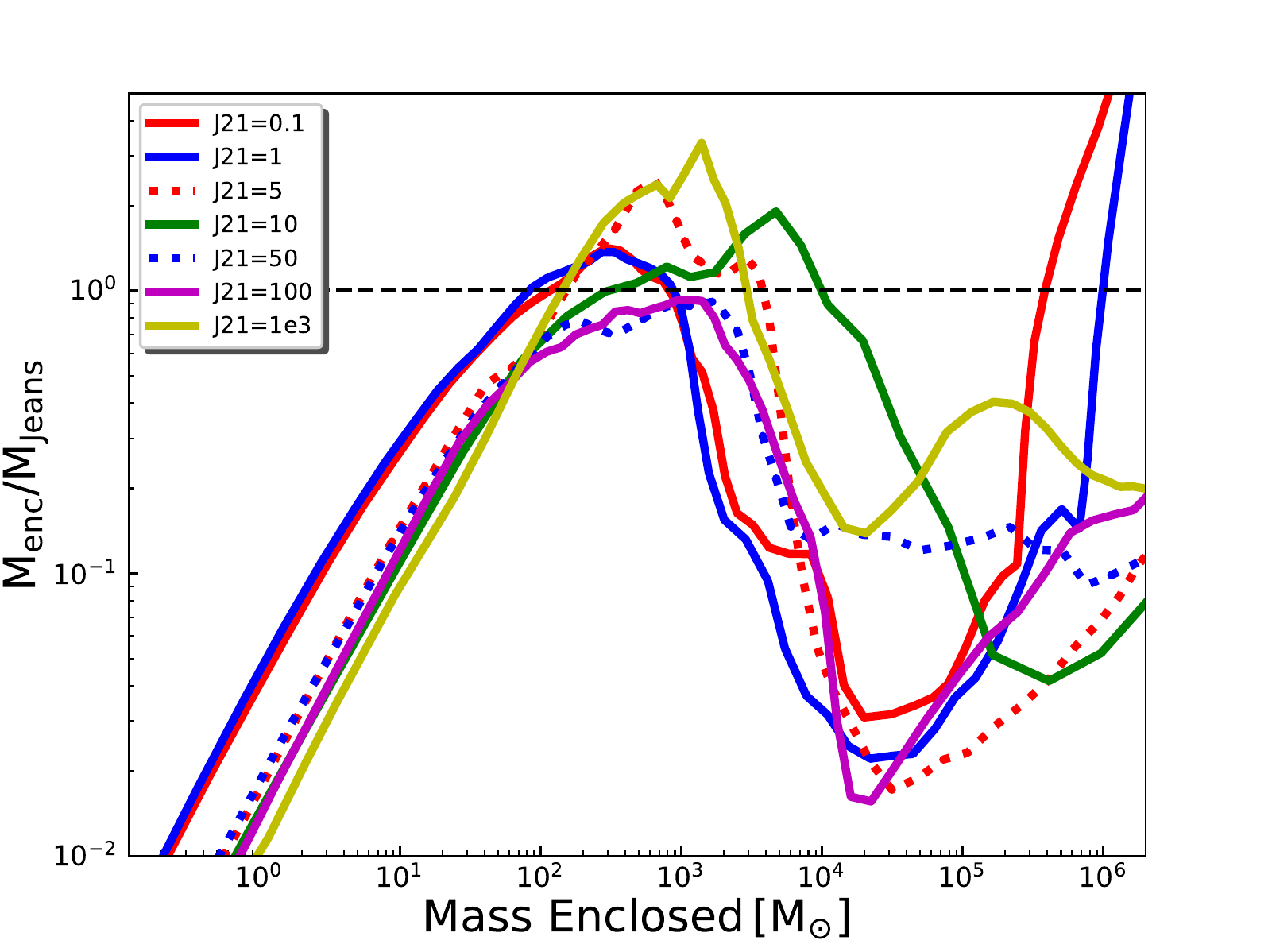}
\end{center}
\caption{The ratio of total enclosed mass to the thermal Jeans mass for a representative halo. The peak value of $\rm M_{enc}/M_{Jeans}$ indicates the gravitationally unstable mass available for star formation.}
\label{fig4}
\end{figure*}

\begin{figure*} 
\begin{center}
\includegraphics[scale=0.8]{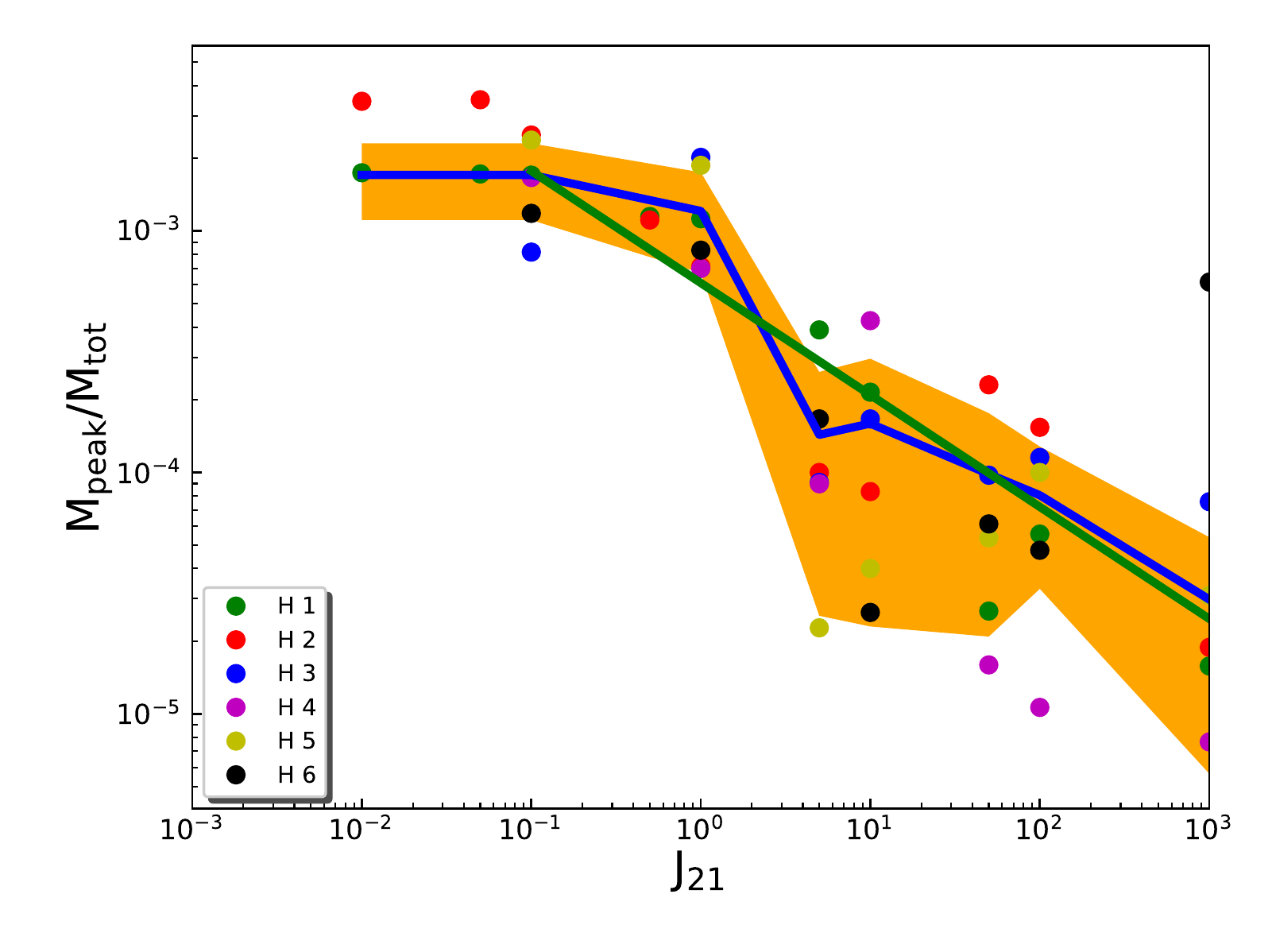}
\end{center}
\caption{This plot shows the ratio of $\rm M_{peak}/M_{tot}$ against the strength of UV radiation field in units of $\rm J_{21}$. The fraction of cold star forming gas decreases by about two orders of magnitude for $\rm J_{21}=1000$ compared to the $\rm J_{21}=0.1$. The colored filled circles represent different halos as mentioned in the legend, H1-H6 stand for halo 1 to halo 6 respectively. The orange shaded region shows the 1-sigma deviation from the mean value. The blue solid line shows the mean value of the ratio and green solid line is fit to the data given in Eq. 2.}
\label{fig5}
\end{figure*}

To estimate the amount of cold dense gas available for star formation in the presence of various UV fields, we plot the ratio of the Jeans mass to the enclosed mass against the total enclosed mass as shown in Fig. \ref{fig4} for a representative case. We compute the enclosed mass by estimating the total gas mass within a sphere of given radius and divide it by the local thermal Jeans mass to assess its stability against gravitational collapse. This ratio monotonically increases with enclosed mass, peaks around a characteristic mass mentioned below and then declines. The peak corresponds to the onset of $\rm H_2$ cooling in the central parsec region of the gas cloud.  We observe this characteristic behaviour in all simulated halos. The $\rm M_{\rm{encl}}/M_{\rm{Jeans}}$ ratio peaks around one and indicates that the gas cloud is subject to gravitational collapse. Based on the location of the peak ratio, we estimate the  cold dense gas mass available for star formation.  In order to compute the fraction of total baryonic mass that will form stars for a given UV radiation field, we plot the ratio of gas mass corresponding to the peak ($M_{\rm{peak}}$) in Fig. \ref{fig4}  to total mass against the strength of radiation flux in Fig. \ref{fig5}.  The typical mass of dense molecular gas in the simulated halos is about $\rm 1000~M_{\odot}$. This coincides with the expected average star formation efficiency of a few percent as seen in the Kennicutt-Schmidt law for local galaxies \citep{Leroy08, Espada19}.

We find that for $\rm J_{21} \leq 1$ the ratio of $\rm M_{peak}/M_{tot}$ is $\rm \sim 10^{-3}$ and almost remains constant. While for larger radiation fluxes $\rm J_{21} > 1$ the ratio of $\rm M_{peak}/M_{tot}$ decreases sharply between $\rm 1 < J_{21} < 5$ and then almost linearly declines with increasing flux with a minimum value of about $\rm 3 \times 10^{-4}$. Some variations from halo to halo are observed due to the different  growth histories, formation redshifts and variation in halo masses but the overall trend is the same for all simulated halos. This suggests that for weaker radiation fluxes ($ \rm J_{21} \leq 1$)  the fraction of star forming gas is $\rm \sim 10^{-3}$ independent of the radiation field while for stronger radiation fields $\rm J_{21} \geq 10$ the fraction of cold gas exponentially declines with $\rm J_{21}$ due to the fact that $\rm H_2$ dissociation becomes more effective. The threshold halo mass for collapse for a given UV radiation flux can be estimated (in units of solar masses) with the following analytical expression:
\begin{equation}
\begin{aligned}
 M_{\rm{th}} =\exp[19.38 - 4.3 \times \exp[-0.16 \times \ln(\Tilde{J})]] ~~ {M}_{\odot} .
\end{aligned}
\end{equation}
Here $\Tilde{J}$ is the value of UV flux in units of $\rm J_{21}$. Similarly the amount of cold dense gas (in units of solar masses) dependent on incident UV flux  $\rm J_{21} \geq 1$ can be well approximated by the following analytic fit:
\begin{equation}
\begin{aligned}
M_{\rm{cold,gas}} = \exp[-0.4637472 \times \ln(\Tilde{J}) - 7.402866] \times M_{\rm{th}}.
\end{aligned}
\end{equation}




\begin{figure*} 
\begin{center}
\includegraphics[scale=0.8]{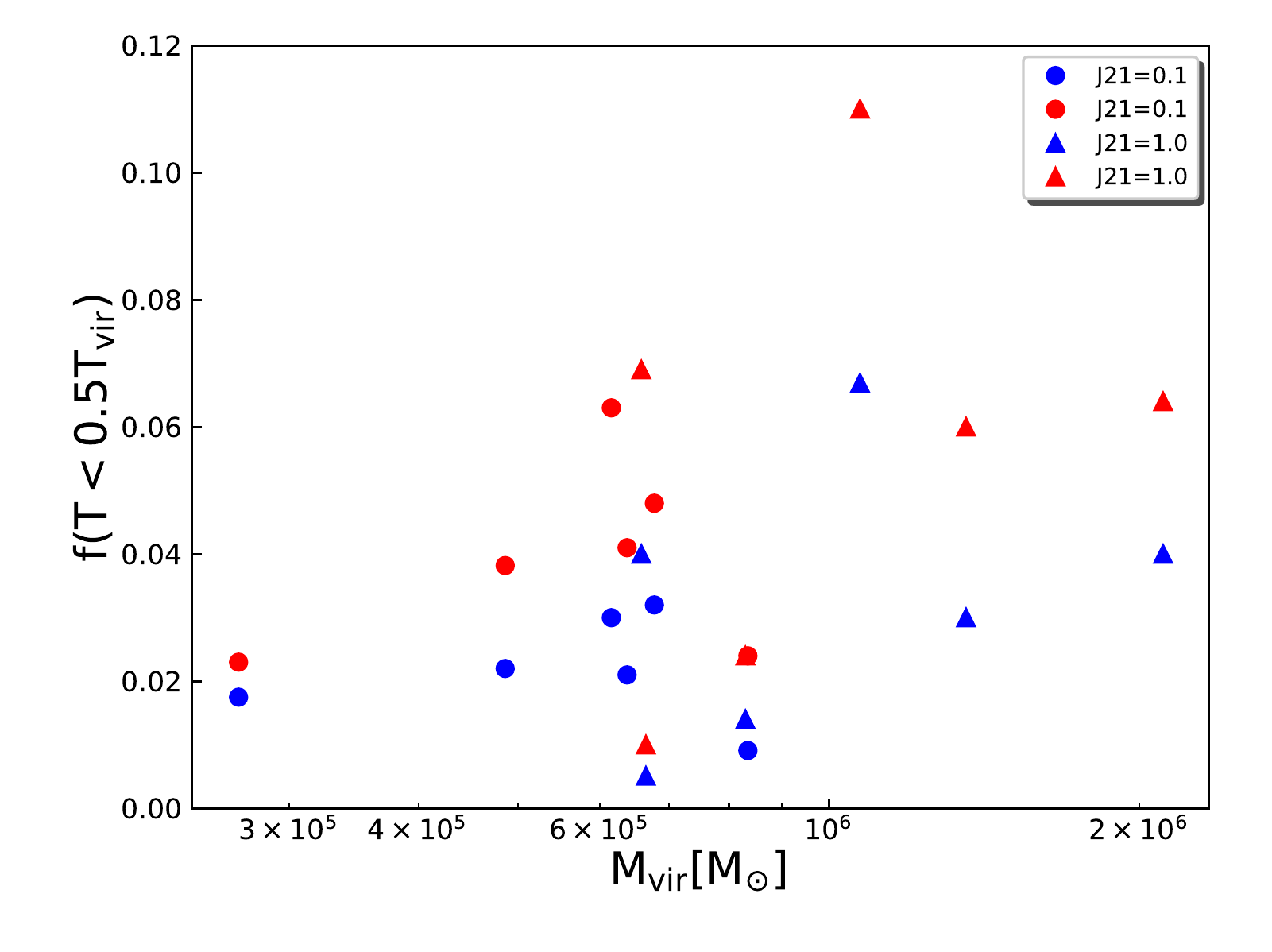}
\end{center}
\caption{Fraction of cold gas against the halo virial mass for $\rm J_{21}=0.1$ and $\rm J_{21}=1$. The red symbols show the fraction of gas cooled by $\rm H_2$ with $\rm T < 0.5 T_{vir}$ and $\rm \rho >$ 1000 times the cosmic mean. The Blue symbols represent the cold dense gas fraction with $\rm T < 0.5 T_{vir}$ and $\rm \rho > 10^{19} ~M_{\odot}/pc^3$.  The plot is same as Fig. 3 of Machacek et al. 2001. }
\label{fig6}
\end{figure*}

\begin{figure*} 
\begin{center}
\includegraphics[scale=1.2]{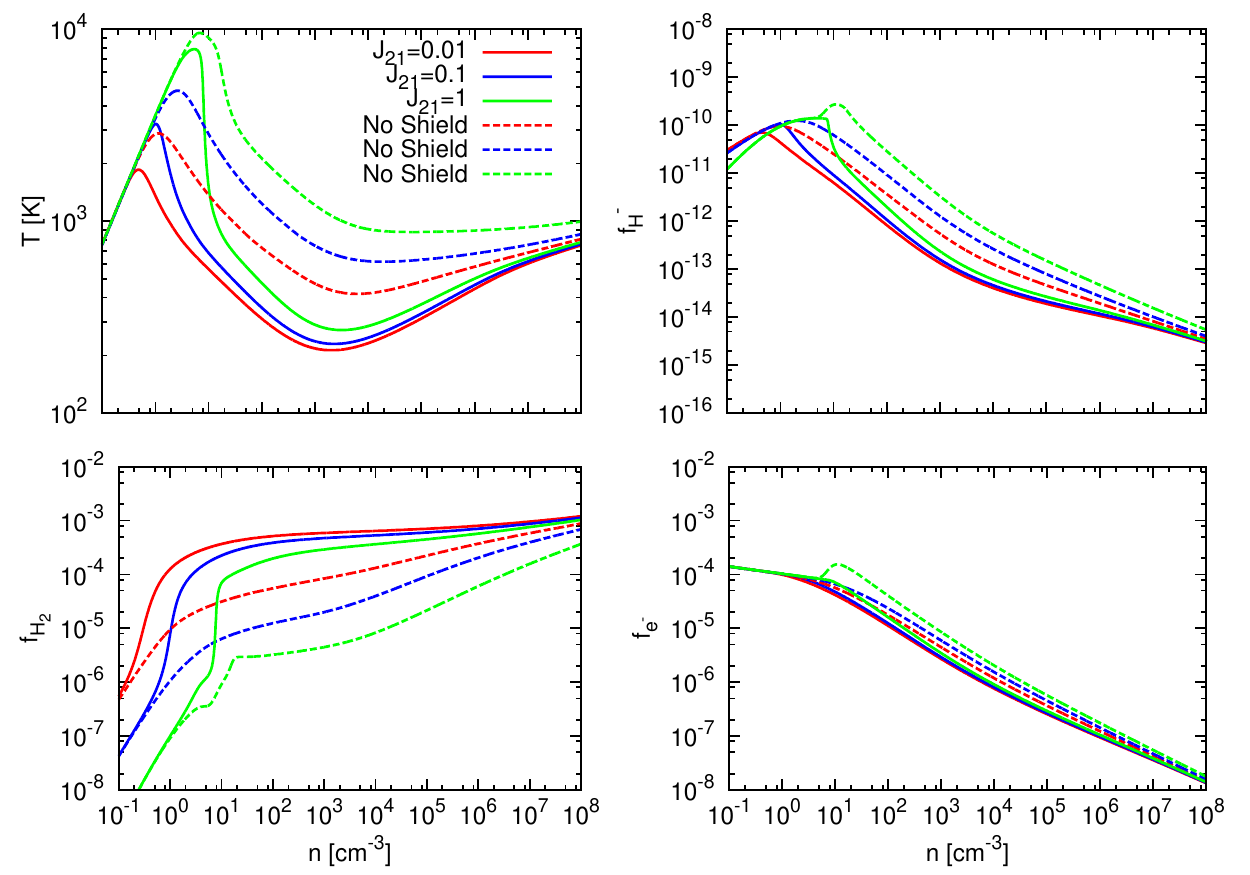}
\end{center}
\caption{Temperature and $\rm H_2$, $\rm H^-$, $\rm e^-$ abundances are plotted against the gas density for $\rm J_{21}= 0.1$ and $\rm J_{21} = 1$ with and without $\rm H_2$ self-shielding.}
\label{fig7}
\end{figure*}

\begin{figure*} 
\begin{center}
\includegraphics[scale=1.2]{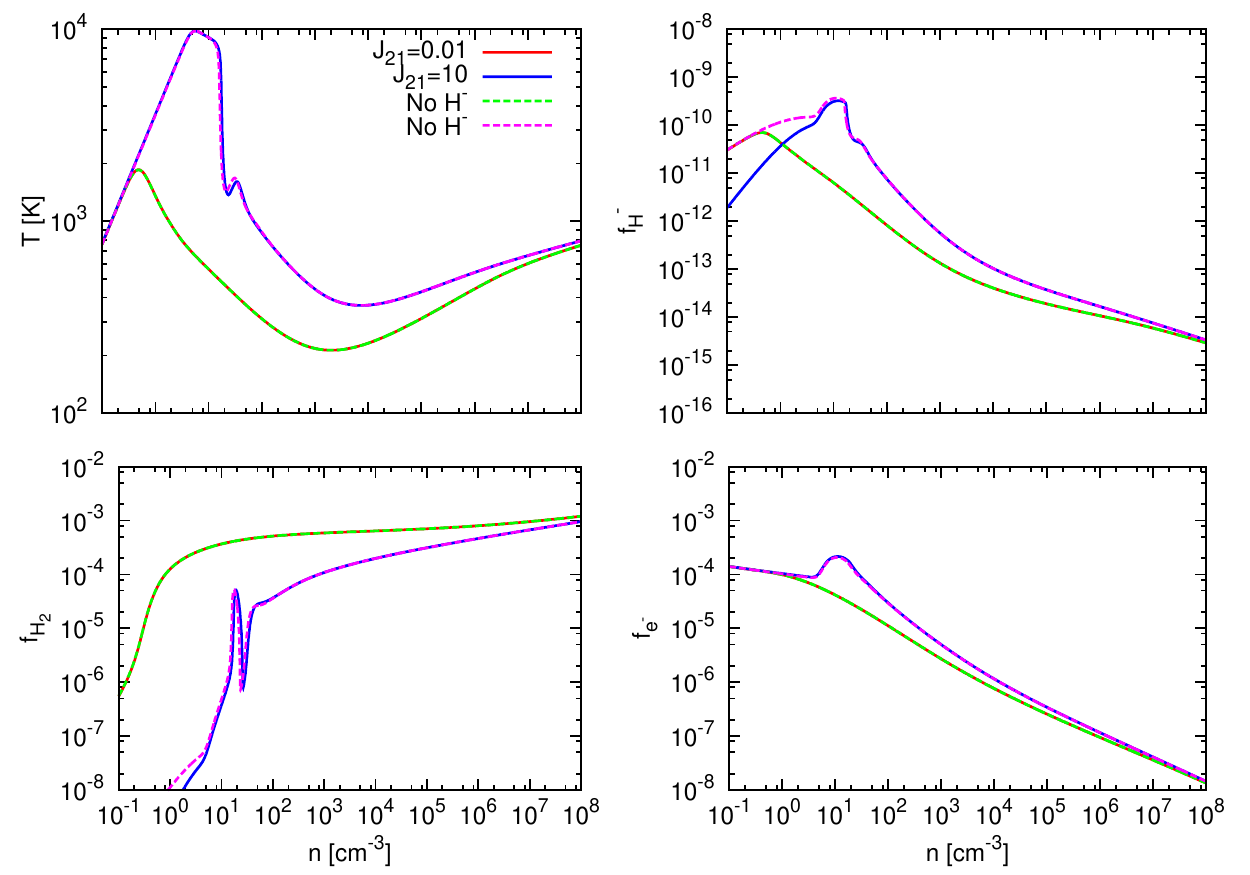}
\end{center}
\caption{Temperature and $\rm H_2$, $\rm H^-$, $\rm e^-$ abundances are plotted against the gas density for $\rm J_{21}= 0.1$ and $\rm J_{21} = 10$ with and without $\rm H^-$ photo-detachment.}
\label{fig8}
\end{figure*}

\begin{figure*} 
\begin{center}
\includegraphics[scale=1.2]{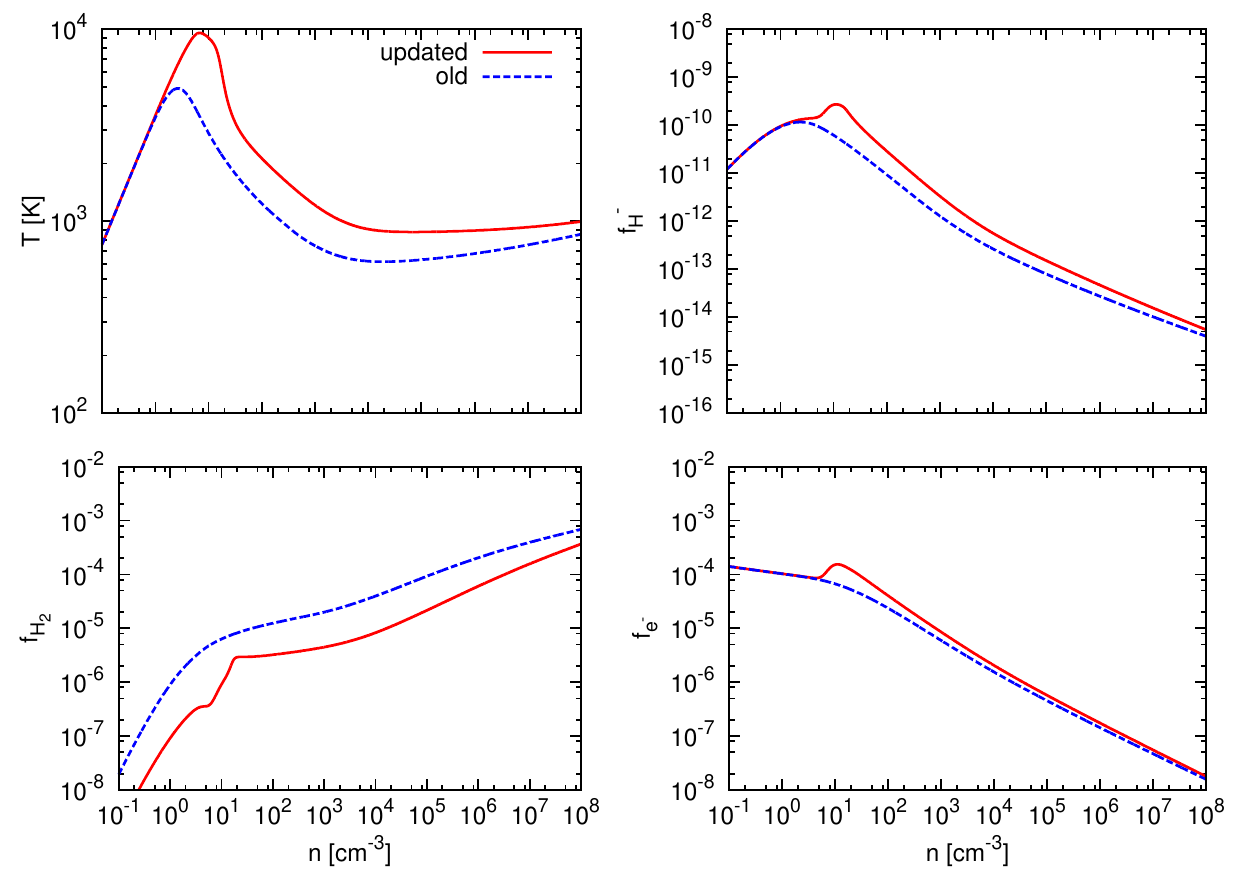}
\end{center}
\caption{Temperature and $\rm H_2$, $\rm H^-$, $\rm e^-$ abundances are plotted against the gas density for updated reaction rates used in this work. They are compared with old reaction rates used in M1.}
\label{fig81}
\end{figure*}

\begin{figure*} 
\begin{center}
\includegraphics[scale=0.8]{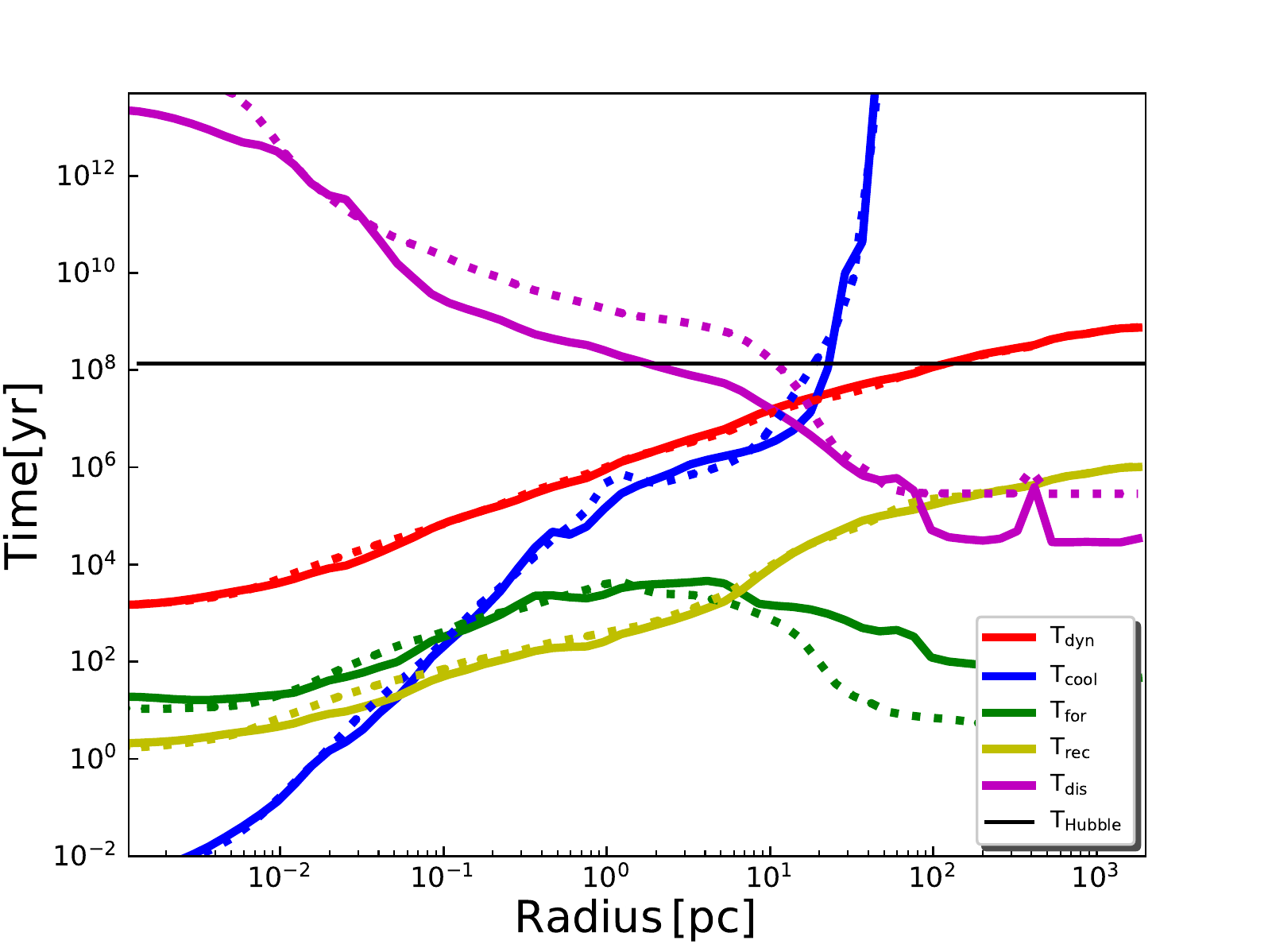}
\end{center}
\caption{Comparison of various timescales relevant for collapse dynamics at $\rm z = 24$ for halo 1 as a representative case. The solid lines for $\rm J_{21}=0.1$ and dashed line for $\rm J_{21}=1$.}
\label{fig9}
\end{figure*}

\subsection{Comparison with previous works}
In order to directly compare our results with previous work of \cite{Machacek01} (hereafter M1), we have adopted the definition presented in M1 and plotted  the $\rm H_2$ cooled and cold dense ($\rm T < 0.5 T_{vir}$ and $\rm \rho > 10^{19} ~M_{\odot}/pc^3$) gas fractions in Figure \ref{fig6}. These fractions are computed in the same way as in Fig. 3 of M1. We find that the $\rm H_2$ cooled gas fraction is slightly higher than the dense gas  and is almost independent of halo mass. Overall, only a few percent of the total gas mass is dense enough to be able to form stars. M1 found that the cold gas fraction logarithmically increases with cloud mass while our findings suggest that the fraction of gas is almost independent of halo mass for $\rm J_{21} \leq 1$, while for stronger radiation fluxes, the fraction of cold dense gas exponentially decreases. These differences are  due to the fact  that M1 ignored the effect of $\rm H_2$ self-shielding as well as photo-detachment of $\rm H^-$ by low energy photons. We find that by considering the  effect of $\rm H_2$ self-shielding the $\rm H_2$ fraction  in the center of the halo is comparable to the no UV case while in M1 $\rm H_2$ is an order of magnitude lower for $\rm J_{21}=1$. Consequently, the fraction of cold dense gas is same for all simulated clouds in our case.

To further quantify the impact of newly added physics, we have performed one-zone calculations to study the impact of $\rm H_2$ self-shielding, updated reaction rates and $\rm H^-$ photo-detachment. The thermal evolution  and  $\rm H_2$, $\rm H^-$ and electron mass fractions  with and without $\rm H_2$ self-shielding against gas density are shown in Figure \ref{fig7}. We find that ignoring self-shielding, $\rm H_2$ direct dissociation becomes more effective and the $\rm H_2$ fraction decreases by about two orders of magnitude depending upon the strength of the incident radiation field. Consequently, the thermal evolution of the gas is strongly impacted and temperatures increase by  an order of magnitude. Neglecting self-shielding also increases the degree of ionization and the $\rm H^-$ abundance. To study the effect of $\rm H^-$ photo-detachment, we run one zone chemical models with and without $\rm H^-$ photo-detachment results of which are shown in Figure \ref{fig8}. We find that the impact of $\rm H^-$ photo-detachment is negligible for $\rm J_{21} \leq 1$. The abundances of $\rm H_2$, $\rm H^-$, $e^-$ and the thermal evolution remain same, while for $\rm J_{21}  \geq 10$ the $\rm H^-$ abundance is decreased by an order of magnitude at low densities but the overall thermal evolution and H$_2$ fractions remain similar. We also find that employing updated reaction rates changes the $\rm H_2$ abundance and thermal evolution by a factor of a few, see Figure \ref{fig81}. Particularly, we note that the most significant contribution comes from the new $\rm H_2$ direct dissociation rate ($\rm k_{diss}$). We have here employed the updated $\rm k_{diss}$ rate from \cite{GloverJappsen2007} and found that the one employed by M1 (originally from \cite{Abel97}) underestimates $\rm H_2$ dissociation. In summary, ignoring $\rm H_2$ self-shielding further increases the threshold $\rm H_2$ cooling mass and reduces the amount of cold dense gas. Therefore, we conclude that among the above mentioned physical processes $\rm H_2$ self-shielding is the most important factor for estimating the cold dense gas in high-redshift galaxies.

To understand the relevance of physical timescales we have estimated the dynamical, cooling, $\rm H_2$ formation, $\rm H_2$ dissociation, recombination and  Hubble timescales. They are plotted in figure \ref{fig9} for the cases of $J_{21}= 0.1$ and 1. Cooling times are shorter than the dynamical time which ensures gas is efficiently cooled by $\rm H_2$ aiding its collapse. The $\rm H_2$ formation, cooling and dynamical timescales are shorter than the Hubble time within the virial radius of the halo.  The $\rm H_2$ dissociation timescale increases within the viral radius due to  $\rm H_2$ self-shielding otherwise  remains constant and shorter than the Hubble time. This suggests that due to $\rm H_2$ self-shielding photo-dissociation equilibrium is not achieved here as compared to M1.  The recombination timescale is also shorter than the dissociation time scale indicating that further $\rm H_2$ formation via H$^-$ channel is suppressed. Overall, the $\rm H_2$ abundance is close to the equilibrium value in our simulations. Without $\rm H_2$ self-shielding  $\rm H_2$ formation timescales become longer and approaches the dissociation time scale. In such situations photo-dissociation equilibrium may be achieved as in M1.

 The maximum resolution in our simulations is about 50 AU in proper units, at-least two orders of magnitude higher than M1. Due to higher resolution in our simulations the central gas cloud is very well resolved and enables us to accurately estimate the dense gas mass in simulated halos. Our findings provide more accurate estimates of cold dense star forming gas in UV irradiated halos.

\section{Conclusions and Discussion}
In this study we have performed cosmological simulations to explore the impact of UV radiation emitted by the first galaxies in regulating star formation in high redshift halos. In total, we have conducted about 50 simulations for six distinct halos by varying the strength of the impinging radiation field  below the Lyman limit from $0.01- 1000$ in units of $\rm J_{21}=10^{-21}~erg/cm^2/Hz/Sr$. In our simulations, we employ a comprehensive chemical model to solve non-equilibrium chemistry of nine primordial species along with cosmological simulations. Particularly we include both photo-detachment of $\rm H^-$, direct photo-dissociation of $\rm H_2$ and self-shielding of $\rm H_2$ as it is the prime coolant for star formation in primordial galaxies.

Our findings suggest that the level of UV radiation needed to suppress $\rm H_2$ formation and delay the gravitational collapse is strongly affected by the ability of gas to self-shield against radiation. This results in the threshold mass for cooling to be  enhanced up to the atomic cooling limit.  For $\rm J_{21} \geq 10$ the collapse of halos is delayed by $\rm \Delta z  \geq 10$ and the halo mass is increased up to a few times $\rm 10^7~M_{\odot}$. On average the collapse is delayed by almost 100 Myrs. We find that the fraction of gas available for star formation is regulated by the UV flux. For  weaker fluxes ($ \rm J_{21} \leq 1$) the fraction of cold star forming gas is about $\rm 10^{-3}$ while for a strong radiation field it exponentially declines with the intensity of the field. The minimum fraction of cold dense gas is about $\rm 10^{-5}$ for $\rm J_{21}=100$ about two order of magnitude lower than for the no UV background cases.

We have simulated halos in different environments  such as  halo 1 and halo 6.  The latter has a quiescent growth history compared to the former and is sitting in  under-dense region in the cosmic web. The same is true for halo 4. As mentioned above halo 6 reaches isothermal collapse for $\rm J_{21}=1000$ and has negligible cold dense gas  contrary to halo 1.  In general,  halos in under-dense regions have lower $\rm H_2$ fraction, slightly higher temperatures and consequently, the amount of cold dense fraction is lower. However, these differences are within 1-sigma deviation from the mean value. The impact of environment becomes much more pronounced in the absence of $\rm H_2$ self-shielding.

In this work we have focused on the impact of  UV radiation but high redshift galaxies may also emit X-rays possibly produced by X-ray binaries, massive stars and even mini-quasars. X-rays  have two competing effects, they heat the gas at low densities  ($\rm  \sim 1 ~cm^{-3}$), enhance the ionization fraction and consequently boost $\rm H_2$ formation ($\rm  \sim 100~ cm^{-3}$). This has been studied in previous work which found that X-rays enhance the amount of cold gas \citep{Kuhlen05,Jeon2012, Inayoshi2012,Hummel15,Latif2015a}. However, \cite{Hummel15} found that X-rays do not change the characteristic mass of stars formed as they get attenuated at high densities. Baryonic acoustic oscillations generate streaming motions which may become supersonic during the cosmic dark ages and influence structure formation \citep{Tselia10}. The impact of streaming motions has been studied via 3D simulations and found to enhance the threshold mass for $\rm H_2$ cooling and delay the collapse depending upon the strength of streaming velocity \citep{Maio2011,Fialkov12,Stacy2012,Latif2014Stream,Schauer19}. Therefore, we expect that strong baryonic streaming motions may further delay star formation for weaker UV radiation fields.


We have employed here a fixed background UV flux  but in reality the UV flux is expected to be time dependent. We have performed one simulation by employing the time variable background UV flux expected  from first galaxies using the estimates  given in \cite{Johnson13}. We found that results are similar to the weaker UV flux cases as the expected background UV flux is a few times $\rm 10^{-2}~ \times ~J_{21}$. We plan on investigating this issue in subsequent simulations.  We have assumed here that UV flux is emitted from first galaxies  dominated by Pop II stars  and their spectra  below the Lyman limit  can be mimicked with $\rm T_{rad} = 2 \times 10^4$ K  as shown by previous studies \citep{Sugimura14, Latif2015a,Agarwal2015}. The simulated UV flux range is assumed to be generated by both the global background as well as local flux emitted from nearby halos.  Based on the cosmic star formation rate density extrapolated from observations  the  fluxes larger than the expected UV background are produced in nearby (single or multiple) halos depending on their star formation efficiency.  The average time for which we have turned on a constant UV flux  is less than 100 Myr in most cases with the exception of $\rm J_{21} \geq 500$ where the flux is assumed constant for 300 Myr.  Depending on the initial mass function,  lifetimes of Pop II stars are expected to be much longer than  the maximum  simulated time here.   While individual galaxies might show fluctuations in their star formation rate on time scales shorter than $\rm \sim100~ Myrs$ studies by \cite{Finlator12}, \cite{Jaacks12}  and \cite{Salmon2015}  suggest that the average star formation rate of galaxies  at $z >10$ are approximately constant over these time scales.  This justifies our choice of keeping the constant UV flux turned on throughout the simulations.

 Moreover, simulated halos with $\rm J_{21}\geq  50$  may get polluted by metals from nearby sources.  Previous studies found that halos located at distance $\rm >$ 10 kpc  remain pristine due to the  preferential ejection of metals into  low density medium  \citep{Ritter2014,Pallottini2015,Habouzit2016a,Latif2016dust,Agarwal17}. Metal enrichment may also be avoided in a synchronized pair of halos where the source halo forms first and the target forms later  \citep{Visbal2014}.  Stronger radiation flux cases with $\rm J_{21} \geq100$ require star formation rates of $\rm \geq ~1~M_{\odot}/yr$.  In such situations, halos may get photo-evaporated from ionizing radiation. \cite{Chon17} performed radiation hydrodynamical simulations  to explore the impact of ionizing radiation in close pairs of halos and found that dense clumps and filaments in the vicinity of source galaxy shield halos from ionizing radiation.  Even in some cases ionizing radiation accelerates the gravitational collapse.  So, we expect halos to avoid photo-evaporation unless they are formed inside HII region.  Given the large delay in the collapse redshifts of halos for stronger radiation fluxes, Pop III star formation in  primordial halos is expected to continue down to $\rm z=10$. \cite{Johnson19} have estimated Lyman-alpha emission from Pop III star forming halos and  found that bright clusters of Pop III stars at $z \sim 10$ may be detectable with the James Webb Space Telescope using the NIRCam survey, which might form in halos with delayed star formation.

\section*{Acknowledgements}

MAL thanks the UAEU for funding via startup grant No. 31S372. 




\bibliography{smbhs.bib}





\bsp	
\label{lastpage}
\end{document}